\documentclass[useAMS,usenatbib    
]{mn2e}    
\usepackage{times}    
\usepackage{graphicx}    
\usepackage{amsmath}    
\usepackage{natbib}    
\usepackage{color,ulem}  
\usepackage{amssymb}         
\bibpunct{(}{)}{;}{a}{}{,}    
         
\title[Constraints from the fundamental plane]{Constraints on the radio loud/radio quiet dichotomy from the fundamental plane} 
   
\author[Garofalo et al.]     
{David Garofalo$^{1}$%\thanks{E-mails:\,david.garofola@gmail.com}    
Matt I. Kim$^{2}$, Damian J. Christian$^{2}$\\
$^{1}$Department of Physics \& Astronomy, Southern Polytechnic State University\\    
$^{2}$Department of Physics \& Astronomy, California State University, Northridge}   
\begin{document}    
    
\date{Accepted . Received ; in original form }    
    
\pagerange{\pageref{firstpage}--\pageref{lastpage}} \pubyear{2011}    
    
\maketitle    
    
%\label{firstpage}    
    
\begin{abstract}   
The fundamental plane for black hole activity constitutes a tight correlation between jet power, X-ray luminosity, and black hole mass.  Under the assumption that a Blandford-Znajek-type mechanism, which relies on black hole spin, contributes non-negligibly to jet production, the sufficiently small scatter in the fundamental plane shows that black hole spin differences of $\mid$$\Delta$a$\mid \sim$1 are not typical among the active galactic nuclei population. If $-$ as it seems $-$ radio loud and radio quiet objects are both faithful to the fundamental plane, models of black hole accretion in which the radio loud/radio quiet dichotomy is based on a spin dichotomy of a$\sim$1/a$\sim$0, respectively, are difficult to reconcile with the observations.   

We show how recent theoretical work based on differences in accretion flow orientation between retrograde and prograde, accommodates a small scatter in the fundamental plane for objects that do have non-negligible differences in black hole spin values.  We also show that the dichotomy in spin between the most radio loud and the most radio quiet involves $\mid$$\Delta$a$\mid \approx$0. And, finally, we show how the picture that produces compatibility with the fundamental plane, also allows one to interpret other otherwise puzzling observations of jets across the mass scale including 1) the recently observed inverse relation between radio and X-rays at higher Eddington ratios in both black hole X-ray binaries as well as active galactic nuclei and 2) the apparent contradiction between jet power and black hole spin observed in X-ray hard and transitory burst states in X-ray binaries.
 
\end{abstract}    
    
    
\section{Introduction}    
      The combination of observations, analytical solutions of the basic equations, and numerical simulations over the past fifty years, have solidified the idea that jet formation is fundamentally an accretion related phenomenon. Observationally, however, the connection between jets and disks is not one-to-one, i.e. accretion does not guarantee that a jet will be present.  While accretion modes that span the entire gamut from sub-Eddington, through thermal accretion, and possibly up through super-Eddington accretion, may all be associated with jets, the association is by no means obvious.  This so-called Ôjet-disk connectionÕ is one of the most pressing problems in high-energy astrophysics (Rawlings \& Saunders 1991; Falcke \& Biermann 1995; Falcke, Malkan \& Biermann 1995; Maraschi \& Tavecchio 2003; Corbel et al. 2004; Fender \& Belloni 2004; Sambruna et al 2006; Chatterjee et al. 2009; Neilsen \& Lee 2009; Evans et al 2010; Garofalo 2013b; Fukumura et al 2014).  
     On galactic scales, the solution to the jet-disk connection is needed to resolve the radio loud/radio quiet dichotomy, the observational fact that only about 20\% of active galaxies have jets (Kellerman et al 1989). While the mode of accretion is in some way likely related not only to the presence, but also to the detailed properties, of the jet, the decades-old spin paradigm for black hole accretion and jet formation requires that powerful jets may only form if the black hole spin is sufficiently large.   The radio quiet AGN in this framework, on the other hand, are predominantly low spinning black holes (Blandford 1990; Wilson \& Colbert 1995; Moderski, Sikora \& Lasota 1998; Sikora et al 2007). However, a variety of observations emerging over the past decade pose serious challenges to the fundamentals of this framework.             

     Two observational facts $-$ the redshift distribution of FRI and FRII radio loud galaxies and quasars, and the radio loud/radio quiet dichotomy at about 20\% $-$ must be addressed by any model for jet production.  Why are FRII quasars on average at higher redshift compared to FRI radio galaxies? And the explanation must be consistent with scale-invariance.  This is a strong constraint with which the spin paradigm continues to struggle (Garofalo 2013a).

     Assuming that broad iron line measurements result from general relativistic effects near the black hole (e.g. Brenneman 2013), the implied distribution of spin in Seyfert galaxies tends to be top-heavy (Brenneman \& Reynolds 2006, 2009; Zoghbi et al 2010; Brenneman et al 2011, Brenneman 2013) which produces tension with the idea that spirals predominantly host low-spinning black holes (King, Pringle \& Hoffmann, 2008). In fact, the high spin values inferred are highly unlikely from pure merger scenarios (Berti \& Volonteri 2008), leaving prolonged accretion as the best explanation. However, differences between chaotic and prolonged accretion are less pronounced at higher mass (i.e. M $>$ 10$^7$ solar masses; e.g. Fanidakis et al 2011) which is where most of the spins have been determined (Reynolds 2013).  While the difficulty in producing high spins via chaotic accretion at higher masses, therefore, is somewhat eased, why are such objects overwhelmingly radio quiet or without powerful jets?  In fact, recent observations point to the existence of powerful jets in a small sample of spiral galaxies (Komossa et al 2006; Foschini 2011).  While this is at odds with the idea that spirals predominantly host low-spinning black holes, it is not a refutation of the basic principles of the spin paradigm.  However, problems with the spin paradigm remain since the observed relation between jet power and Eddington ratios in gamma-narrow line Seyfert 1 ($\Gamma$-NLS1) galaxies is sufficiently different from that observed in flat spectrum radio quasars (FSRQ-Abdo et al 2009; Ghisellini et al 2010; Foschini 2011), which scale-invariance within the spin paradigm would tend to make equivalent (i.e. the Eddington ratios are expected to be the same in the spin paradigm).  In fact, while typical AGN black hole mass uncertainties of a factor of 3 may distort the vertical shift between $\Gamma$-NLS1s and FSRQs more than in decreasing the horizontal shift in radio loudness between the two populations on the Eddington ratio vs radio loudness plane (Figure 1 Garofalo 2013a), the differences remain.  

     Additionally, in X-ray binaries, the X-ray bright hard state jet power appears not to correlate with black hole spin (Fender et al 2010), which is at odds not only with the simple Blandford-Znajek-based principles of the spin paradigm, but also with recent numerical simulations whose jet power spin-dependence is even steeper (Tchekhovskoy et al 2010).  However, recent observation-based work suggests a steep spin-dependence for ballistic jets (Narayan \& McClintock 2012), which, under the assumption that the physical mechanism for producing the jet is the same, is contradictory (but, see Russell et al 2013 who strongly dispute this correlation based on the limited choice of sources in Narayan \& McClintock reaching Eddington accretion values). 

     Further constraints come from the observed relation between jet power, black hole mass, and X-ray luminosity, known as the fundamental plane for black hole accretion (Merloni, Heinz \& Di Matteo 2003; Falcke 2004; Laor \& Behar 2008; Gultekin 2009).  The small scatter in this relation argues that internal parameters that may influence jet power (most notably black hole spin), are not dominant.  While the fundamental plane may not involve a unique track, but two separate paths for each of the radio loud and radio quiet objects (i.e. Li et al 2008), the differences in spin values associated with the two paths tend to be sharp. In other words, small scatter in any track(s) on the fundamental plane implies little variation in additional parameters such as black hole spin.  While the possibility of separate tracks may well emerge from discontinuities in other parameters (such as those that govern the jet suppression in galactic black hole binaries), two sharply divided spin populations (one low, one high) is not the most natural outcome of black hole evolution, which would be a continuous distribution of spins. This expectation of a continuous distribution of black hole spins is difficult to reconcile with the fundamental plane and a spin-based radio loud/radio quiet dichotomy. In fact, both a single-track fundamental plane as well as correlations between radio and optical are difficult to reconcile with $\mid$$\Delta$a$\mid \sim$1 for the radio loud/radio quiet dichotomy (van Velzen \& Falcke 2013).  In other words, while the spin paradigm suggests that the radio loud/radio quiet dichotomy involves $\mid$$\Delta$a$\mid \sim$1, a one-track fundamental plane relation as well as the radio-optical correlations in FRII quasars, argue that $\mid$$\Delta$a$\mid\sim$0, or that differences in spin somehow do not matter, i.e. that the spin paradigm is not the correct foundation for jet production. While other parameters may not be ruled out in explaining the fundamental plane relation(s), black hole spin does not appear to be an appealing strategy.

Our goal in this paper is to argue that the observations are not incompatible or mutually contradictory, and that the fundamentals of the so-called gap paradigm which appear to resolve one apparent problem, also ease the tension with the others.  We will press for the idea that both families of extreme AGN (i.e. the most radio loud and the most radio quiet) have similar values of black hole spin, so that the radio loud/radio quiet dichotomy is not spin-based, and that a large class of jet-producing AGN do so in a way that changes little with differences in spin. This does not imply, however, that jet production is not spin-dependent, simply that it is not spin-dependent as prescribed in the spin paradigm. We show how the simple ideas at the heart of our accretion disc orientation-based dichotomy accommodate $\mid$$\Delta$a$\mid \sim$0 for the radio loud/radio quiet division, a by-product of which is a resolution to the apparent contradictions in the observations of jets across the mass scale. In Section 1 we illustrate the basic elements of our orientation-based theoretical framework that are needed to interpret the above mentioned observations; in Section 2 we discuss how the phenomenology explains the observations and present our fundamental plane-like plot. Lastly, in Section 3 we conclude.

\subsection{The gap paradigm}
      The gap paradigm for black hole accretion, jet formation and collimation, and disk winds, constitutes a scale-free phenomenological framework for modeling black hole accretion in X-ray binaries and AGN (Garofalo, Evans \& Sambruna 2010).  While superficially very little appears to be needed compared to the standard spin paradigm, the crucial element of the orientation of the accretion flow in either retrograde or prograde configurations, leads to fundamental differences.  In a retrograde accretion configuration, both the Blandford-Znajek  (Blandford \& Znajek 1977) and Blandford-Payne mechanisms (Blandford \& Payne 1982) are maximized (Garofalo 2009b). This is due to the fact that circular orbits are stable further away from the black hole in the retrograde configuration so the gap region between the inner edge of the accretion disk and the black hole is larger. And, a larger gap region produces a larger magnetic flux accumulation on the black hole due to the rapid infall onto the black hole of the plasma, with the magnetic flux being effectively frozen into the infalling gas (Garofalo 2009a). This larger flux influences not only the Blandford-Znajek effect via the increased strength of the field on the black hole, but also the Blandford-Payne disk jet by the extra bending of the disk field produced by the black hole threading flux bundle (Garofalo 2009b).  Hence the most powerful and most collimated jets occur in higher-spinning black holes surrounded by retrograde accretion.  

In the prograde regime, on the other hand, radiatively efficient accretion at higher prograde spin is associated with a relatively larger disk efficiency and correspondingly larger disk wind power (Kuncic \& Bicknell 2004, 2007) due to the fact that the inner edge of the accretion disk is closer to the black hole, which results in a greater amount of energy being reprocessed into the accretion disk.  If larger disk winds are effective at smothering the jet, as observations suggest (Done et al 2007; Neilsen \& Lee 2009; Ponti et al 2012), these ideas imply a spin-dependence to the jet quenching or jet-suppression ability of a radiatively efficient disk.  This is the scale-free mechanism adopted in the gap paradigm for claiming that high-spinning, prograde black hole systems in radiatively efficient accretion states, are radio quiet or weak jet producers.  Hence, depending on the orientation of the accretion material, the high-spinning black hole is either a powerful jet producer or a very weak one.  But it is important to emphasize how this competition between the jet and the disk occurs in the context of radiatively efficient accretion states.  In low Eddington, radiatively inefficient flows, instead, the jet suppression ability of the disk drops due to the weakness of disk winds. Therefore, X-ray binaries with high spinning prograde black holes produce jets in the X-ray hard states but weak ones in the X-ray soft states.   In GRS 1915+105, for example, the jet suppression mechanism comes into play as the system transitions to the soft state where the disc wind dominates the dynamics (Neilsen \& Lee 2009).  It is interesting to notice that determinations of black hole spin value for this source span the upper prograde range from 0.7 (Middleton et al 2006) to close to unity (McClintock et al 2006).  In line with the gap paradigm, the lower value of spin implies that the jet suppression ability of the soft state is comparatively weaker than the ability of a radiatively efficient accretion disc surrounding a higher prograde-spinning black hole. In this light it would be interesting to produce a scale-invariant analysis of the relative strength of the jet suppression ability as a function of black hole spin from the observations, which could shed light on the threshold value of black hole spin associated with jet suppression.  In short, there is an inverse relation between the effectiveness of the jet and that of the disk wind in radiatively efficient states. 

While the radio loud/radio quiet dichotomy in the gap paradigm is due to the difference in accretion orientation around a fast spinning black hole, the framework also offers a natural explanation for the observation that the fraction of radio loud objects over radio quiet objects is less than 1. Assuming for simplicity that mergers lead to accretion disk orientations that 50\% of the time are retrograde and 50\% of the time are prograde (Dotti et al 2010), the fraction of retrograde accretion disks that evolve via prolonged accretion into high prograde spins while remaining in a radiatively efficient accretion state, become radio quiet quasars/AGN. If this occurs at the Eddington limit, the high prograde regime is reached in less than about 10$^8$ years.  Because jet suppression fails in radiatively inefficient accretion states, the originally radio loud quasars whose accretion states do not remain in their radiatively efficient mode throughout the prograde spin-up phase, evolve into low excitation FRI radio galaxies due to their now radiatively inefficient accretion mode (Garofalo, Evans \& Sambruna 2010).  Hence, there is a natural mechanism associated with prolonged accretion, which turns a fraction of the originally radio loud quasars into radio quiet quasars.  No mechanism, on the other hand, exists in the paradigm which accomplishes the inverse, i.e. that takes radio quiet quasars and turns them into radio loud quasars via prolonged accretion.  The quantitative details of the fraction of radio loud quasars to radio quiet quasars depends on the assumptions one makes about mergers.  Another mechanism, however, further shifts the ratio of radio loud to radio quiet to lower values.  This comes from the recognition that retrograde accretion is more unstable as the mass ratio of the black hole to the accreting matter is smaller (Perego et al 2009).  Hence, for a given accretion mass, smaller black holes will tend to flip to a prograde accretion configuration, thereby further decreasing the ratio of radio loud to radio quiet objects.  While we need to understand better both the nature and probability of retrograde occurrence in postmerger systems, as well as the instability of retrograde accretion, before we can tackle the observed quantitative value of about 20\% for the radio loud to radio quiet ratio in the gap paradigm, there nonetheless is a mechanism favoring a greater density of radio quiet quasars over radio loud quasars as the redshift decreases.  Notice also that an explanation exists in this framework for why the FRII quasar density peaks at higher redshift (about z = 2), as well as to why the radio loud population tends to have the more massive black holes (Floyd et al 2013). Within the context of this scenario, we string together a variety of different observations, arguing for both a logically consistent and simple phenomenological picture.

\section[]{The Observations}    
% 2.1 
 \subsection[]{X-ray binaries}
Fender, Gallo \& Russell (2010) have searched for correlations between values of black hole spin in X-ray binaries and proxies for jet luminosity and have concluded that the spin dependence is weak for all accretion states but in particular for the hard X-ray state.  While this is problematic for the spin paradigm, the gap paradigm also models such X-ray hard states as radiatively inefficient, sub-Eddington accretion onto prograde black holes; the crucial difference, however, is that the jet power vs spin dependence in the gap paradigm is flatter (Figure 3 of Garofalo, Evans \& Sambruna 2010).  While this flatter spin dependence for jets in the prograde regime lessens the tension with the radio and infrared observations of Fender et al (2010), Figure 3 of Garofalo et al (2010) still displays a clear spin dependence of jet power (i.e. while the jet power increases less steeply as the spin increases in the gap paradigm, the trend of larger jet power for larger black hole spin still exists).  In other words, the jet power vs spin in the gap paradigm is still incompatible with the observations of Fender et al (2010). However, the jet power increase with prograde spin is almost certainly a feature of the force-free solution of Garofalo, Evans \& Sambruna (2010). Neglecting the inertia of the accreting plasma (the hallmark of a force-free magnetosphere), makes it easier to break the centrifugal barrier that rapidly rotating black holes create, thereby allowing the jet power to increase more than otherwise as the spin increases in the prograde direction.  In other words, we are suggesting that a more realistic set of equations that include the inertia of the plasma in the magnetosphere, would lead to a further flattening of the spin dependence in the prograde regime, thereby producing a better match between the basic structure of the gap paradigm and the observations of Fender et al (2010).  In fact, this physics would operate in the retrograde regime as well, thereby making the accumulation of magnetic flux onto the black hole more difficult at high spins, irrespective of the prograde or retrograde orientation of the disc.  In other words, while the overall jet power would be larger in the retrograde regime, plasma inertia would flatten the spin dependence of jet power both in the high prograde and high retrograde spin cases.  This requires both inertia and resistivity, a tall order and the subject of future work.  In short, dynamically relevant inertia should operate to weaken the steepness of jet power dependence on black hole spin at high spin.

     What about the ballistic jets?  For such states, Narayan \& McClintock (2012) find a clear spin dependence of jet power.  Here, we assume the validity of this correlation but note again that this is disputed (Russell et al 2013). And, again, the validity of this correlation appears to be at odds with Fender et al (2010) for the X-ray hard state jets.  However, according to the prescription of the gap paradigm, the ballistic jet requires the transitory disk wind for collimation; hence, the collimation of the jet will depend on the power of that wind which in turn depends on disk efficiency.  But, as discussed above, the disk efficiency for a radiatively efficient disk depends on the value of black hole spin, larger for larger prograde black hole spin. Hence, the ballistic jet should display a spin-dependence in its jet, unlike the jet in the X-ray hard state. The X-ray hard state jet, in fact, is associated with a hot, radiatively inefficient accretion flow, so its jet is not collimated by a transitory disk wind, and no additional spin dependence appears beyond that of Figure 3 in Garofalo et al (2010).  It may be useful to point out that Figure 3 in Garofalo et al (2010) does not incorporate the jet suppression mechanism.  A greater suppression of the jet occurs at higher prograde spin due to the location of the innermost stable circular orbit and the reprocessing of the energy into the disc leading to stronger winds (Kuncic \& Bicknell 2004, 2007). While gap paradigm phenomenology requires a collimation of X-ray hard state jets by a Blandford-Payne disk jet, it is flatter than in the retrograde case because the Blandford-Payne jet is weakest in the high prograde regime (Garofalo 2009b).  In a scale-invariant sense, therefore, the model predicts that bright X-ray hard-state jets in X-ray binaries are less collimated than jets in FRII quasars and FRII radio galaxies and that ballistic jets in X-ray binaries are not the small-scale equivalent to FRII quasar/FRII radio galaxy jets.  According to gap paradigm phenomenology, in fact, the ballistic jet in X-ray binaries does not have an AGN counterpart due to the absence of AGN evolution from radiatively inefficient accretion to radiatively efficient accretion.  This is a basic feature of the time evolution of jetted AGN in the gap paradigm (Garofalo, Evans \& Sambruna 2010). Hence, we have shown how both the findings of Fender et al (2010) and those of Narayan \& McClintock (2012) can be qualitatively reconciled within one phenomenology.  We now turn to the issues raised by the fundamental plane for black hole activity.

%2.2 
 \subsection[]{The fundamental plane}
The fundamental plane for black hole activity constitutes a correlation between proxies for jet power, and X-ray luminosity and black hole mass (Merloni et al 2003; Falcke et al 2004; Gultekin et al 2009). The tightness of the correlation argues that factors internal to the jet engine such as black hole spin cannot be fundamentally relevant.  In other words, the apparent small scatter in the correlation leaves little space for varying quantities that are not X-ray luminosity and black hole mass, and still getting a significant difference in jet power.  But the validity of this produces a glaring contradiction within the spin paradigm.  In this framework, in fact, the radio loud/radio quiet dichotomy involves high-spinning black holes in the radio loud population and low-spinning black holes in the radio quiet population.  But, such large variation in black hole spin between the two populations (i.e. of order unity) would affect the fundamental plane. But this does not seem to be the case even if we include powerful radio quasars. In fact, very recent work shows that a tight correlation exists between jet power and bolometric luminosity for a large sample of FRII quasars at redshift of z $\sim$ 1 (van Velzen \& Falcke 2013).  
 
 %\section{Results}    
 
\begin{figure*}    
\includegraphics[width=105mm]{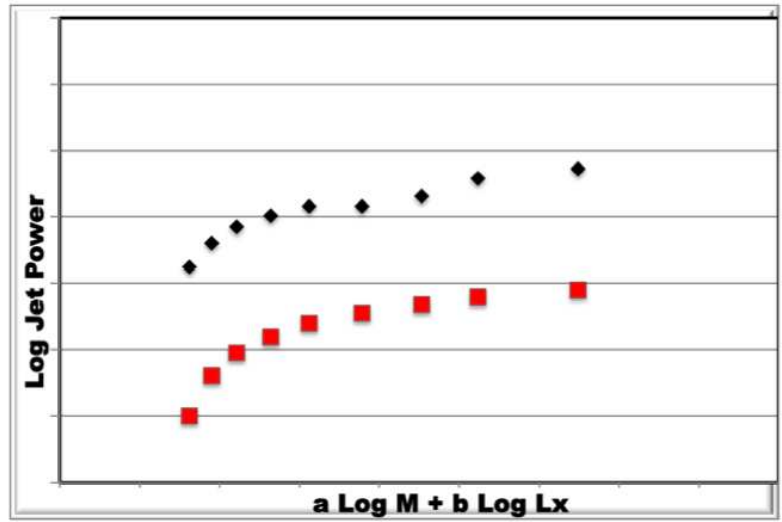}  
\caption{A mock fundamental plane for the full range of prograde spinning black  holes in the standard spin paradigm (red) and the gap paradigm (black). The vertical axis represents the logarithm of jet power while the horizontal axis plots a logM+b logLx with a and b arbitrary constants of order unity.  We avoid a scale since there is an overall normalization ambiguity due to unknown magnetic field strengths.  The two order of magnitude difference in jet power, however, correctly captures differences in jet power between the two paradigms. The X-ray luminosity is assumed to come from accretion and scaled with the mass of the black hole.  The jet power in the standard spin paradigm scales with the square of the spin of the black hole while in the gap paradigm it is modulated by the $\alpha$$\beta$$^2$ term (Garofalo et al 2010).  
The emphasis should be on the scatter between the objects in the two paradigms.  While the gap paradigm has jet powers that are constrained within two orders of magnitude despite a wide range in spins and the force-free solution, the standard spin paradigm jet powers span a range that exceeds two decades.  This constitutes a lower limit on the scatter since numerical simulations generate steeper spin dependence at high prograde spins.
}
\label{cmd}    
\end{figure*}    

Such observations suggest that the radio loud/radio quiet dichotomy is characterized by $\mid$$\Delta$a$\mid \approx$0. Our goal in this section is to show how the gap paradigm prescribes $\mid$$\Delta$a$\mid \approx$0 for the most extreme radio loud and radio quiet objects, and that a smaller scatter results for LINERS, Seyferts, other low luminosity AGN, FRI radio galaxies, NLS1 and the recently discovered $\Gamma$-NLS1, compared to the spin paradigm, despite $\mid$$\Delta$a$\mid \neq$0 for these classes of AGN. In the gap paradigm, in fact, both of the extremes of AGN - the powerful radio loud quasars as well as the radio quiet quasars - are characterized by high-spinning black holes, the difference being the retrograde vs prograde nature of the radiatively efficient accretion flow (Sambruna et al 2009, 2011; Garofalo et al 2010; Ballo et al 2011).  In fact, as outlined in section 1, a maximal black hole spin in a retrograde orientation of the disk constitutes the most effective of jet-producing conditions, maximizing both the Blandford-Znajek and Blandford-Payne mechanisms; while a prograde accretion configuration produces the most effective jet-quenching or jet-suppression conditions, with strongest disk winds due to the large disk efficiency.  

     While $\mid$$\Delta$a$\mid \approx$0 between the most radio loud and most radio quiet objects is achieved by imposing scale-invariance and adopting jet suppression in soft states in X-ray binaries (Neilsen \& Lee 2009; Ponti et al 2012), other AGNs such as Seyferts, LINERS, other low luminosity AGN, FRI objects, NLS1 and $\Gamma$-NLS1, produce a smaller scatter in the gap paradigm compared to the spin paradigm despite the possibility of a wide range in prograde spin values for these objects.  To show this, we produce a fundamental plane-like relationship for these AGN using the different prescriptions for jet power adopted in the gap and spin paradigms.  We plot jet power vs a log M + b log Lx, where M and Lx are black hole mass and X-ray luminosity in dimensionless form and a and b are order unity constants. While we assume the X-ray luminosity originates in the accretion disk and we scale it according to black hole mass, the horizontal axis will not produce differences between the paradigms.   What matters is the fact that the two paradigms associate different jet powers as a function of spin (larger in the gap model) and different dependencies on spin, the latter being the focus here.  The difference between the two functions is completely captured by the $\alpha\beta^2$ term in equation (20) of Garofalo, Evans \& Sambruna (2010), for prograde values.  In addition to these differences as a function of spin, the two paradigms also differ in the range of spin values prescribed for each class of AGN.  The gap paradigm allows a greater range of spin for most of the classes discussed here. For example, NLS1 objects span the intermediate to high prograde spin range, the $\Gamma$-NLS1 objects span the low to intermediate prograde spin range, LINERS, low luminosity AGN, and FRI radio galaxies span almost the entire prograde spin range.  The FRII radio quasars and FRII radio galaxies, on the other hand, are of course modeled as retrograde accreting black holes. In the spin paradigm, on the other hand, any jetted AGN class tends to live in a narrow upper prograde spin range (note the recent observation of an FRI object as an intermediate prograde spin system with a$\approx$0.6, which produces tension with the spin paradigm Ð Doeleman et al 2012).  While variation in black hole mass over many orders of magnitude will produce a plot representing the fundamental planes for each paradigm, this is not the simplest way of capturing and describing the differences.  We therefore fix the black hole mass and allow the spin value to span the entire prograde range.  As a result, our plot involves jet power vs X-ray luminosity, with the alogM term an overall constant.  Hence, Figure 1 is not the fundamental plane for either paradigm.  Its purpose is to highlight the scatter that each paradigm produces under the assumption that a large range in spin is possible. We have allowed the spin range to span most of the prograde values from 0.1 to maximal spin.  The point is to explore which framework produces less scatter for such a large range in spin values. While the differences are not large, the spin paradigm produces a scatter in jet power that spans more than two orders of magnitude for a given black hole mass while the gap paradigm generates a scatter that is less than two decades in jet power (recall that this is in addition to the aforementioned near-zero scatter in spin between the most radio loud and most radio quiet objects in the gap paradigm compared to maximum scatter in the spin paradigm Ð i.e. $\mid$$\Delta$a$\mid \sim$1 in the latter for the radio loud/quiet dichotomy). By spin paradigm, note, we are referring to models that incorporate the simple Blandford-Znajek spin-squared dependence of jet power on black hole spin.  Given, as previously pointed out, that numerical simulations suggest much steeper dependence on black hole spin at high prograde spin, the scatter they produce is greater than shown in Figure 1, so our result should be thought of as a lower limit on the scatter for the spin paradigm.  For the gap paradigm, on the other hand, the scatter observed in Figure 1 is likely an upper limit due to the aforementioned assumed force-free nature of the black hole magnetosphere.  Note that since the gap paradigm prescribes a steeper spin dependence in the retrograde regime, FRII quasars should populate the fundamental plane with a larger scatter compared to radio loud AGN with FRI jet morphology, radio quiet quasars, LINERs, NLS1 etc., which in the gap paradigm are all modeled as prograde accreting systems (see Table 1 for list of object class and model prescription).  However, as pointed out in section 2.1, the centrifugal barrier of a non force-free magnetosphere would tend to flatten out the jet power at highest retrograde spin as well so the scatter there would be less than that produced by the $\alpha\beta^2$ term.  While the arguments tend to rule out black hole spin, there are mechanisms in the gap paradigm  (such as that behind the radio loud/radio quiet dichotomy and that for jet suppression in radiatively efficient vs inefficient accretion modes) that are to some degree discontinuous processes, a fact that may be useful in exploring the existence of multi-track fundamental planes (Li et al 2008; Corbel et al 2013; Dong et al 2014).  As our goal is to address black hole spin, we do not explore this further here.

     In summary, in this section we have proposed a picture whereby on the one hand, the extremes of the radio loud/radio quiet dichotomy are not distinguished by spin, and, on the other for objects that do differ in spin, have shown that they do not differ much in jet power.  In other words, largest differences in jet power occur where the spin differences are zero (but accretion orientations are opposite), and smaller differences in jet power occur where differences in spin can be large.  It is these basic features of the gap paradigm that make it appealing in our search for a simple explanation to the existence of a fundamental plane-type relationship for black holes across the mass scale.  But it is important to recognize that the zero difference in spin between the most radio loud and most radio quiet objects is the result of assuming scale-invariance in the context of jet suppression. Hence, the jet suppression mechanism connected to the transient ballistic jet in microquasars via the disk wind, is also at the heart of our discussion concerning the radio loud/quiet dichotomy in AGN so that a common phenomenology applies to different observations across the entire mass scale.

%2.3 
\subsection{An inverse relation between radio and X-ray flux}
While AGN generally appear to live on the fundamental plane, they may cross that plane in different ways. Recent work suggests an inverse relation between radio and X-ray flux for objects sharing Eddington ratios of 1$-$10\% as observed in NGC 4051, 3C 120 and two X-ray binaries (King et al 2011). This inverse relation, again, is interpreted in the gap paradigm as resulting from a competition between jets and disks, which depends on the size of the gap region.  As the region decreases, the disk efficiency increases and the disk wind begins to dominate over the jet, suppressing the jet as the system crosses the threshold prograde spin value (Garofalo 2013b).  When the gap region, on the other hand, is larger, the disk efficiency is smaller and disk winds are weaker while jets are most effective.  It is important to note that the time evolution of black hole X-ray binary states is not based on the change in the gap region since the spin value does not change appreciably on the observed timescales. But this scenario is valid at non-negligible Eddington ratios. As the radiative efficiency drops below about 0.01 Eddington, jet power and proxies for disk efficiency increase in tandem with increase in prograde black hole spin, as the sub-Eddington disk struggles to compete with, and to suppress, the jet (Garofalo et al 2010).  Therefore, for NGC 4051, which has a measured spin that is high although possibly not maximal (Brenneman 2007; Patrick et al 2011; Brenneman 2013), and an Eddington luminosity of only about 3\% ($\pm$2\% at one sigma due to uncertainties in black hole mass - Peterson et al 2004; Denney et al 2009), we have a transitional object. Since this object is near the demarcation line between radiatively efficient and inefficient $-$ usually thought to be near 2\% Eddington (e.g. Esin et al 1997; Dunn et al 2010) although we note that X-ray hard states may exist at higher Eddington ratios $-$ the jet suppression should be weaker than for higher Eddington values, which observationally, in fact, appears to be the case.  Alternatively, the weakness of the jet suppression mechanism could also be due in part to the possibility that the spin is near but below the threshold value for jet suppression (i.e. a $>$ 0.67 in Brenneman 2007 while a $<$ 0.94 in Patrick et al 2011).  Converging on a spin value for NGC 4051 promises interesting constraints. 

3C 120, on the other hand, appears to be radiating at almost one third of its Eddington value (Woo \& Urry 2002; with a range spanning 22\% $-$ 63\% including the uncertainty in black hole mass $-$ Pozo Nunez et al 2012), which, according to a superficial approach to our picture would seem to imply that jet suppression should be effective.  Observationally, however, the jet suppression appears to be as weak in this object as it is for NGC 4051, so the theoretical framework requires that the spin value approach the threshold value for jet suppression, which is intermediate prograde (Garofalo 2013).  In other words, the spin would be slightly below this threshold value. In fact, this object appears to be a rare high excitation spiral galaxy with an FRI jet morphology (Ogle et al 2004; Kataoka et al 2007; Hardcastle, Evans \& Croston 2009; Abdo et al 2010), so it fits in the gap paradigm as a prograde accreting AGN. Recent measurements of spin for 3C 120 have spanned a wide range from retrograde (Cowperthwaite \& Reynolds 2012) up to high prograde (Lohfink et al 2013) so this should be an interesting object to focus on, hoping that observations can converge on a more stable spin measurement.  

     Cignus X-1 and GRS 1915+105 $-$ two X-ray binaries $-$ also transition to an inverse relation between radio and X-rays once their Eddington ratios are near the 2\% value and the systems enter the Ôhigh softÕ state (King et al 2011).  While the emphasis of our jet suppression mechanism has been on the size of the gap region and therefore on the value of black hole spin, the wind power in a radiatively efficient Shakura \& Sunyaev disc, depends on the accretion rate as well.  Therefore, different accretion rates should also produce differing degrees of jet suppression. This would be the framework within which to explore differing ballistic jet strengths at different epochs.   We do not explore this further here. The bottom line here is that our simple scale-invariant mechanism for jet suppression in radiatively efficient accretion states (higher Eddington ratios) requires an inverse relation between jet power (radio as a proxy) and thermal disk signatures (X-rays as a proxy), and a direct relation between jet power and disk signatures in radiatively inefficient (lower Eddington ratios) accretion states.
 
 \begin{table*}    
\begin{minipage}{110mm}    
\begin{center}    
\caption{Class of accreting black hole vs. the accretion prescription in the spin and gap paradigms. Since the spin paradigm is grounded in the notion that a geometrically thick disk is needed to produce a jet, the quasar-like objects that are also associated with jets need to be radiatively inefficient in the inner regions, which is why high $\dot{m}$ disks are postulated there.  While the prescription in general is the same for the black hole X-ray binary class, there are significant differences, of which, the most important is that the spin paradigm requires high prograde spin for the objects that produce powerful jets while the gap paradigm does not. Note that the different classes are not necessarily mutually exclusive (i.e. FSRQ generally belong to the High excitation FRII radio galaxy class). ADAF=Advection Dominated Accretion Flow; $\dot{m}$ = accretion rate.}     
\label{table1}    
%\begin{tabular}{@{}cc@{}c@{}c@{}c@{}cc@{}c@{}c@{}c@{}cccccr@{}c@{}l@{}}  
  \begin{tabular}{lll}
  \hline    
 {\bf Class} & {\bf Spin Paradigm}  & {\bf Gap Paradigm} \\ 
\hline    
\hline    
\\    
%Class                        Spin paradigm                                           Gap paradigm 
 High excitation FRII radio galaxy    &    high prograde/high $\dot{m}$ ADAF      &         retrograde/thin disk \\
FSRQ                                              &             high prograde/high $\dot{m}$ ADAF   &          retrograde/thin disk\\
Low excitation FRII radio galaxy       &  high prograde /ADAF                              &    retrograde/ADAF \\
High excitation FRI quasar              &   high prograde/high $\dot{m}$ ADAF           &  intermediate prograde/thin disk \\
Low excitation FRI radio galaxy     &  high prograde/ADAF                           &        prograde/ADAF \\
Radio quiet quasar                       &           near zero spin/thin disk                &             high prograde/thin disk  \\
$\Gamma$-NLS1                          &             high prograde/high $\dot{m}$ ADAF         &    intermediate prograde/thin disk\\
Radio quiet NLS1     &                               near zero spin/thin disk               &              high prograde/thin disk\\
LINERs                                 &                       near zero spin/ADAF-like     &                      prograde/ADAF-like \\
Black hole X-ray binaries    &                  prograde/ADAF-thin disk cycle     &          prograde/ADAF-thin disk cycle\\
\hline        
\hline    
\end{tabular}     
\end{center}     
\end{minipage}    
\end{table*} 
 
 \subsection{On retrograde accretion in black hole X-ray binaries}
Recent work suggests that retrograde accretion may do a better job of explaining jet power during the transitory burst state of the BHXRB in M31 (Middleton, Miller-Jones, \& Fender 2014; Middleton et al 2013). While these authors agree that from a theoretical perspective it seems difficult to envision retrograde accretion commonly occurring in BHXRBs, the good fit to a retrograde accretion model suggests otherwise. But the assumption of retrograde accretion produces consequences.  We discuss them here and suggest that the poor fit to the Blandford-Znajek mechanism for prograde black holes may originate from the assumption of a lack of spin dependence in the black hole flux accumulation inherent in the standard Blandford-Znajek effect.
 
     Let us first relax our concerns about the formation of retrograde accretion in BHXRBs in order to explore the implications and difficulties of this from the perspective of the gap paradigm. We are forced to the following picture and restrictions.  First, the bright X-ray hard state jets that form in retrograde BHXRBs would be different from the bright X-ray hard state jets that form in prograde BHXRBs (assuming both prograde and retrograde occurrence in BHXRBs) to the same degree that low excitation FRII radio galaxies are different from low excitation FRI radio galaxies.  In other words, everything else being equal, the collimation and acceleration of the X-ray hard state jets in retrograde configurations would be greater and this would probably be observable in BHXRBs.  Secondly, and perhaps more fundamentally, the large gap regions of retrograde configurations are the reason for both the powerful jet and the relative weakness of the disk wind.  This $-$ as discussed and applied in previous sections $-$ is the foundation behind the relative weakness of jet suppression in retrograde accreting systems and the reason why FRII quasars in the gap paradigm accrete in a way that makes them large scale equivalents to soft states in BHXRBs despite the presence of the jet. The upshot of these ideas is that modeling BHXRBs as retrograde accreting systems implies weakness of jet suppression, which leads us to the following question:  Does the accreting black hole in M31 (XMMU J004243.6+412519) never experience a high-soft state where the jet turns off?  In fact, jets that fail to turn off are a feature of comparatively weak disk winds associated both with retrograde systems and with prograde systems whose spin values are not large (Garofalo, Evans \& Sambruna 2010; Garofalo 2013b).  

     So what do we have to say about the good fit in Middleton et al (2014) for the retrograde assumption?  Our understanding is that Middleton et al (2014) have produced a fit to the standard Blandford-Znajek effect, and not to the spin-dependent-Blandford-Znajek mechanism of Garofalo (2009a), and more importantly, not to the spin-dependent Blandford-Znajek/Blandford-Payne combination explored in Garofalo, Evans \& Sambruna (2010).  Unlike the standard Blandford-Znajek mechanism, Figure 3 of Garofalo, Evans \& Sambruna (2010) shows how a prograde-spinning black hole with spin a = 0.1 experiences an enhanced jet power in the gap paradigm of a factor of 181 due to the $\alpha\beta^2$ factor. This means that in the context of the gap paradigm, a black hole system whose spin is 0.1 will produce a jet power that is 1.81 times as large as that produced by a maximally spinning black hole in the standard Blandford-Znajek effect without the flux enhancement. Therefore, for values of spin that range from maximal spin or just less than a $\approx$ 1,  to values of spin just below a = 0.1, the gap paradigm prescribes jet powers that are larger than the standard Blandford-Znajek effect even for maximal spin. Hence, we suggest exploring whether a fit using the spin dependent factor $\alpha\beta^2$ of Garofalo, Evans \& Sambruna (2010) for jet powers in the prograde regime might produce better compatibility.  Of course, it could be that retrograde accretion occurs in BHXRBs outside of the restrictions of the gap paradigm.  However, there is no scale-invariant mechanism that incorporates such a scenario. 

\section{Conclusions}    
The gap paradigm for black hole accretion and jet formation is pivoted around three types of physical characteristics: black hole spin (high or low), state of accretion (radiatively efficient or inefficient) and orientation of the accretion flow (retrograde or prograde). The possible combinations of these features produce a phenomenological framework for interpreting the radio loud/radio quiet dichotomy in a way that makes it compatible with observations of small scatter for the fundamental plane as well as for observations of radio vs optical.  We have also explored this phenomenology for objects across the mass scale that lie on the fundamental plane but appear to cross it in different ways depending on the radiative efficiency or Eddington ratio of the accretion flow.  In particular, we suggested that the inverse relation between radio and X-ray luminosity in X-ray binaries, Seyferts and quasars, also hinges on the same ideas that resolve the radio loud/radio quiet dichotomy.  And finally, we argued how this phenomenology was relevant in resolving the apparent contradiction between jet power and black hole spin in transient vs X-ray hard state jets in X-ray binaries.

\section*{Acknowledgments}   
We thank an anonymous referee for suggested improvements to the manuscript. MK and DC thank the CSUN Department of Physics and Astronomy for their support of this project.
    
%\bibliographystyle{mn2e} 
%\bibliography{biblio} 
\medskip
\medskip
\noindent {\bf REFERENCES}\\
Abdo, A.A., et al, 2010, ApJ, 720, 912\\
Ballo, L. et al 2011, MNRAS, 418, 2367\\
Berti, E. \& Volonteri, M. 2008, ApJ, 684, 822
Blandford R.D. Courvoisiser and M.Mayor Eds., p. 161, Springer, Berlin, 1990\\
Blandford, R.D. \& Znajek, R. 1977, MNRAS, 179, 433\\
Blandford, R.D. \& Payne, D. 1982, MNRAS, 199, 883 \\
Brenneman L.W., 2013, Springer Briefs in Astronomy, 978-1-7770-9\\
Brenneman L.W., et al 2013, MNRAS, 429, 2662\\
Brenneman, L.W., 2011, ApJ, 736, 103\\
Brenneman, L.W., Reynolds, C.S., 2009, ApJ, 702, 1367\\
Brenneman, L.W. 2007, PhDT, 38B\\
Brenneman, L. \& Reynolds, C.S., 2006, ApJ, 652, 1028\\
Chatterjee, R. et al 2009, ApJ, 704, 1689\\
Corbel, S. et al., 2004, ApJ, 617, 1272\\
Corbel, S. et al 2013, MNRAS, 428, 2500\\
Cowperthwaite, P.S. \& Reynolds, C.S., ApJ, 752, L21\\
Denney, K.D., et al 2009, ApJ, 702, 1353\\
Done, C., Gierlinski, M. \& Kubota, A. 2007, A\&Arv, 15, 1\\
Dong, Ai-Jun et al 2014, arXiv: 1404, 5317\\
Dotti, M. et al 2010, ASPC, 427, 19
Dunn, R.J.H., Fender, R.P., Kording, E.G., Belloni, T. \& Cabanac, C., MNRAS, 2010, 403, 61\\
Evans, D.A. et al 2010, ApJ, 710, 859\\
Esin, A, McClintock, J., Narayan, R., 1997, ApJ, 489, 865\\
Falcke, H., Biermann, P.L., 1995, A\&A, 293, 665\\
Falcke H., Malkan, M.A., Biermann, P.L., 1995, A\&A, 298, 375\\
Falcke H., et al 2004, A\&A, 414, 895\\
Fender R.P, Gallo E., Russell D., 2010, MNRAS, 406, 1425\\
Fender, R. \& Belloni, T. 2004, ARA\&A, 42, 317\\
Floyd, D.J.E, Dunlop, J.S. and Kukula, M.J. 2013, MNRAS, 429, 2\\
Foschini, L. 2011, Res. Astron Astrophys., 11, 1266\\
Fukumura, K. et al 2014, ApJ, 780, 120\\
Garofalo, D., Evans D.A., Sambruna R.M., 2010, MNRAS, 406, 975\\
Garofalo, D., 2013, MNRAS, 434, 3196(a)\\
Garofalo, D., 2013, Adv. Astron., 213105(b)\\
Garofalo, D., 2009, ApJ, 699, 400(a)\\
Garofalo, D., 2009, ApJ, 699, L52(b)\\
Ghisellini et al 2010, MNRAS, 402, 497\\
Gultekin et al 2009, ApJ, 706, 404\\
Hardcastle, M. J., Evans, D.A., \& Croston, J.H, 2009, MNRAS, 396, 1929\\
Kataoka, J. et al, 2007, PASJ, 59, 279\\
Kellerman, K.I. et al 1989, AJ, 98, 1195\\
King A.L. et al, 2011, ApJ, 729, 1\\
King, A.R., Pringle, J.E. \& Hoffman, J.A., 2008, MNRAS, 385, 1621 \\
Komossa, S. et al 2006, AJ, 132, 531\\
Kuncic, Z. \& Bicknell, G.V., 2004, ApJ, 616, 669\\
Kuncic, Z \& Bicknell, G.V., 2007, Ap\&SS, 311, 127\\
Laor, A. \& Behar, E., 2008, MNRAS, 390, 847\\
Lohfink, A. M.,et al. 2012, ApJ, 772, 83 \\
Maraschi, L. \& Tavecchio, F. 2003, ApJ, 593, 667\\
Merloni, A., Heinz,  S. \& Di Matteo, T., 2003, MNRAS, 345, 1057\\
Middleton, M., Miller-Jones, J. \& Fender, R., 2014, MNRAS, in press\\
Middleton, M. et al 2013, Nature, 493, 187\\
Moderski R., Sikora M., Lasota J.P, 1998, MNRAS, 301, 142 \\
Narayan, R. \& McClintock, J. 2012, MNRAS, 419, L69\\
Neilsen, J. \& Lee, J.C., 2009, Nature, 458, 481\\
Ogle, P.M. et al 2004, AAS, 36, 766\\
Patrick, A.R. et al 2011, MNRAS, 416, 2725
Perego, A., Dotti, M, \& Volonteri, M., 2009, MNRAS, 399, 2249\\
Peterson, B.M., et al 2004, ApJ, 613, 682\\
Ponti et al 2012, MNRAS, 422, L11\\
Pozo Nunez, F. et al. 2012, A\&A, 545, A84\\
Rawlings, S. \& Saunders, R., 1991, Nature, 349, 138\\
Reynolds, C.S., 2013, CQGra, 30, 4004\\
Sambruna, R.M. et al 2006, ApJ, 652, 146\\
Sambruna, R.M. et al 2009, ApJ, 700, 1473\\
Sambruna, R.M. et al 2011, ApJ, 734, 105\\
Sikora M., Stawarz L., Lasota J-P, 2007, ApJ, 658, 815\\
Tchekhovskoy, A., Narayan, R., McKinney, J. 2010, Apj, 711, 50\\
van Valzen, S. \& Falcke, H. 2013, A\&A, 557, 7\\
Wilson A.S., Colbert E.J.M., 1995, ApJ, 438, 89\\
Woo, J.H, \& Urry, C.M., 2002, ApJ, 579, 530\\
Zoghbi, A. et al 2010, MNRAS, 401, 2419\\
   
\label{lastpage}    
\end{document}